\documentclass[12pt,pre,pas,showpacs,reprint]{revtex4-1}
\usepackage{graphicx}
\usepackage{geometry}
\usepackage{mathrsfs}
\usepackage{subfigure}
\usepackage{amssymb}
\usepackage{indentfirst}
\usepackage{color}
\usepackage{cancel}
\usepackage{amsmath}
\usepackage{hyperref}
\usepackage[]{appendix}
\usepackage[]{color}
\usepackage{float}
\begin{document}
\title{\LARGE\textbf{Hyperuniformity of generalized random organization models}}
\author{Zheng Ma}
\affiliation{Department of Physics, Princeton University}
\affiliation{Princeton, New Jersey 08544, USA}

\author{Salvatore Torquato}
\email{torquato@electron.princeton.edu} 
\affiliation{Department of Chemistry, 
Department of Physics, Princeton Institute for the Science and Technology of Materials,\\
and Program in Applied and Computational Mathematics, Princeton University,
Princeton, New Jersey 08544, USA}

\begin{abstract}
Studies of random organization models of monodisperse (i.e., identical) spherical particles have shown that a hyperuniform state is achievable when the system goes through an absorbing phase transition to a critical state. Here we investigate to what extent hyperuniformity is preserved when the model is generalized to particles with a size distribution and/or nonspherical shapes. We begin by examining binary disks in two dimensions and demonstrate that their critical states are hyperuniform as two-phase media, but not hyperuniform nor multihyperuniform as point patterns formed by the particle centroids. We further confirm the generality of our findings by studying particles with a continuous size distribution. Finally, to study the effect of rotational degrees of freedom, we extend our model to noncircular particles, namely, hard rectangles with various aspect ratios, including the hard-needle limit. Although these systems exhibit only short-range orientational order, hyperuniformity is still preserved. Our analysis reveals that the redistribution of the ``mass" of the particles rather than the particle centroids is central to this dynamical process. The consideration of the ``active volume fraction" of generalized random organization models may help to resolve which universality class they belong to and hence may lead to a deeper theoretical understanding of absorbing-state models. Our results suggest that general particle systems subject to random organization can be a robust way to fabricate a wide class of hyperuniform states of matter by tuning the structures via different particle-size and -shape distributions. This in turn potentially enables the creation of multifunctional hyperuniform materials with desirable optical, transport and mechanical properties.
\end{abstract}
\date{}
\maketitle

\section{Introduction}
Periodically driven colloidal suspensions have been observed to have a phase transition at the onset of irreversible dynamics \cite{pine2005chaos, corte2008random}. The physics of this absorbing-phase transition \cite{hinrichsen2000non, henkel2008non} was successfully captured by a ``random organization" model \cite{corte2008random}. At around the same time, the hyperuniformity concept, which describes the unusual suppression of number fluctuations of point configurations at large length scales, came to the fore \cite{torquato2003local}. Hyperuniform point configurations possess a structure factor $S(\mathbf k)$ that goes to zero as the wave number $k \equiv |{\mathbf k}|$ vanishes, i.e.,
\begin{equation} \label{auto1}
\lim_{|\mathbf k|\rightarrow 0} S(\mathbf k)=0.
\end{equation}
 A hyperuniform system in $d$-dimensional space $\mathbb{R}^d$ is poised at a special critical point \cite{torquato2003local} in which the direct correlation function, defined via the Ornstein-Zernike relation \cite{hansen1990theory}, is long-ranged, which is the diametric opposite behavior of traditional critical points in which the total correlation is long-ranged. Since 2003, many different
types of disordered hyperuniform systems, including equilibrium and nonequilibrium varieties, have been
identified; see the recent review \cite{torquato2018hyperuniform}.
\begin{figure*}[]
\centering
\subfigure[]{
\includegraphics[width=6cm, height=6cm]{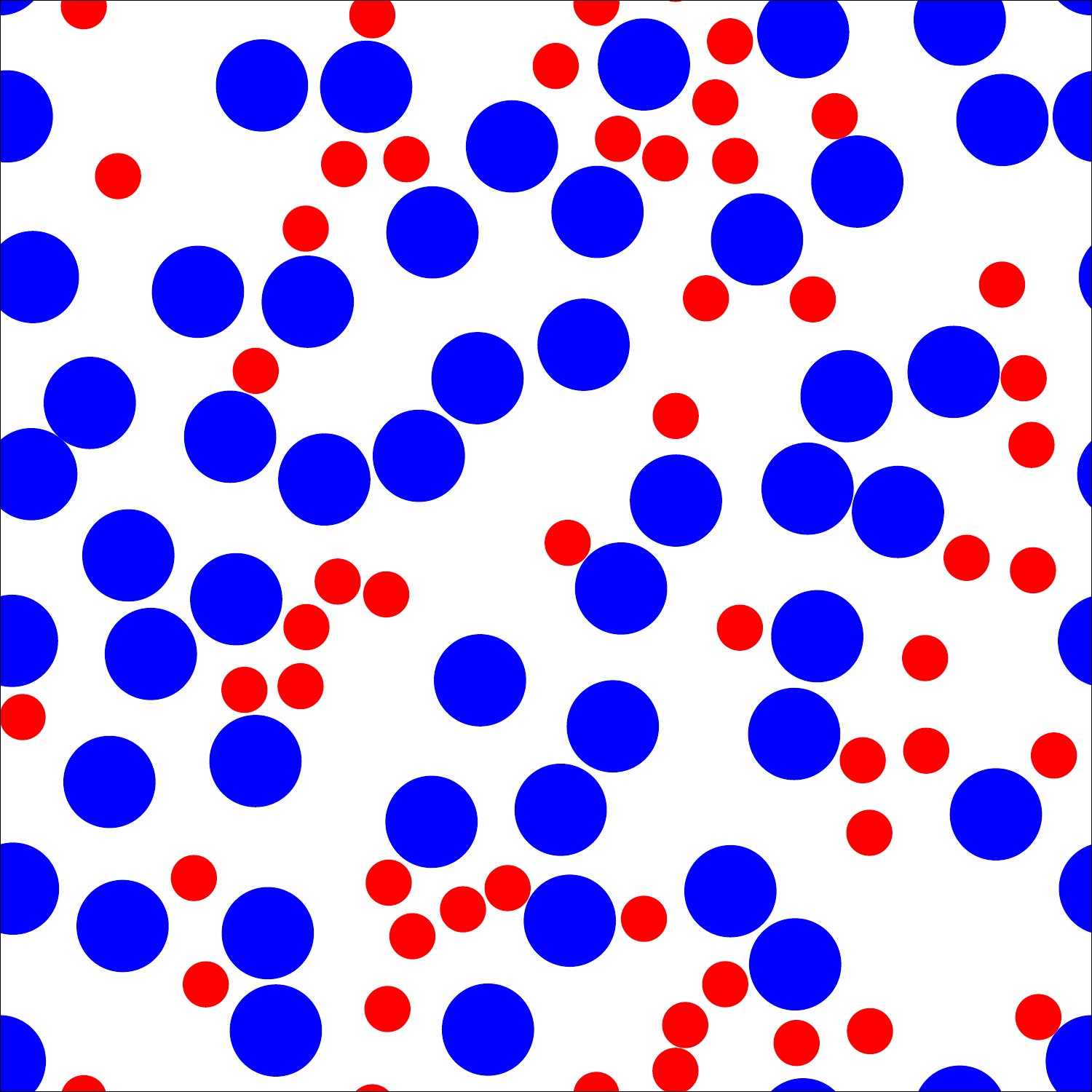}
}
\subfigure[]{
\includegraphics[width=8cm, height=6cm]{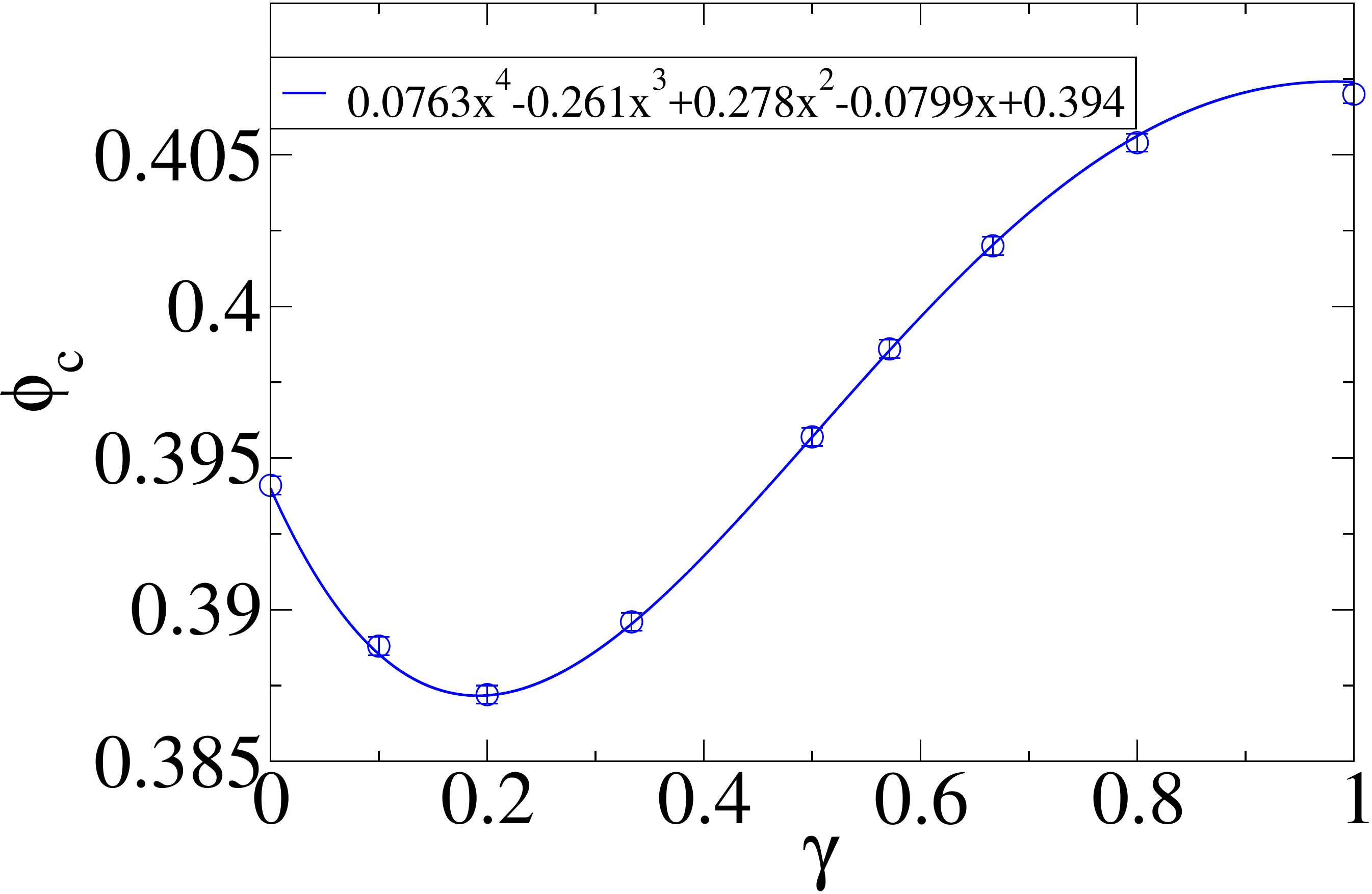}
}
\caption{(a) A representative image of a configuration of a 2D binary disk system in an absorbing state with the small-to-large particle size ratio $\gamma=0.5$ and the relative number fraction of small disks $x=0.5$. (b) Critical volume fraction $\phi_c$ as a function of $\gamma$ with fixed $x= 0.5$. The smooth curve is fit well by a polynomial of degree 4 (blue curve). Interestingly, it is seen that there is a local minimum at $\gamma \approx 0.2$.}
\label{fig: binary}
\end{figure*}
\indent It was only recently that Hexner and Levine showed that the critical absorbing states associated with random organization models are actually hyperuniform \cite{hexner2015hyperuniformity}. They fall in a special class of hyperuniform point configurations called class III \cite{torquato2018hyperuniform}, where the structure factor scales as a power $S(k) \sim k^{\alpha_{_N}}$ with $0<\alpha_{_N}<1$ as $k \rightarrow 0$, and the local number variance scales as a power-law $R^{d-\alpha_{_N}}$.\\
\indent Many variants of such models and systems have been studied numerically \cite{hexner2017enhanced, tjhung2015hyperuniform, dickman2015particle, goldfriend2017screening, wang2018hyperuniformity}. Moreover, experimentally, such protocols suggest that hyperuniform states of matter can be made in a self-organized fashion \cite{weijs2017mixing, weijs2015emergent, zito2015nanoscale}. However, previous numerical studies have focused on models that are based on monodisperse (i.e., identical) spherical particles, which is an idealization that may not be achievable nor tunable in practice. Interestingly, it is not known whether multicomponent systems subjected to random-organization dynamics preserves hyperuniformity, and, if so, how it is preserved. If it is preserved, would the multicomponent system also be ``multihyperuniform" \cite{jiao2014avian, lomba2018disordered} (each species of which the system consists is hyperuniform by its own) at the critical point? Should we consider the system as a two-phase medium, and apply the corresponding generalization of hyperuniformity associated with volume-fraction fluctuations \cite{torquato2016hyperuniformity}? Hyperuniformity is defined for a two-phase medium in terms of the spectral density $\tilde\chi_{_V}(\mathbf k)$ (the Fourier transform of the autocovariance of the phase indicator function \cite{torquato2013random}), which approaches zero as the wave number vanishes \cite{zachary2009hyperuniformity} 
\begin{equation} \label{auto2}
\lim_{|\mathbf k|\rightarrow 0} \tilde\chi_{_V}(\mathbf k)=0.
\end{equation}
This implies that the infinite-wavelength volume-fraction fluctuations vanish identically. Equivalently, the hyperuniformity condition states that the local volume-fraction variance $\sigma_{_V}^2(R)$ decreases more rapidly than $R^{-d}$ for large $R$ \cite{torquato2016hyperuniformity, zachary2009hyperuniformity}. Another unexplored question is whether a system maintains hyperuniformity that consists of nonspherical (noncircular) particles, in which case rotational degrees of freedom are introduced.\\ 
\indent This paper investigates all of these extensions of random organization models and their consequences. In Sec. \ref{sim}, we briefly describe the simulation procedure that we employ. In Sec. \ref{disks}, we investigate the hyperuniformity of the critical absorbing states of two-dimensional models of binary mixtures of circular disks as well as disks with a size distribution.
In Sec. \ref{non}, we carry out a similar study for hard rectangles of various aspect ratios, including the needle-like limit.
Finally, in Sec. \ref{discuss}, we discuss our results and their consequences, including preliminary results concerning the universality class of random organization models.

\section{Simulation Procedure}
\label{sim}

\indent Throughout the paper, we consider the isotropic version of the 
random organization model and use the volume fraction $\phi$ as a control parameter \cite{tjhung2016criticality}. Each system starts from a collection of randomly placed particles. At a given instant of time $t$, any pair of particles that overlap with one another are considered ``active" and are given a random kick in the next time step, while nonoverlapping particles are considered ``inactive" and will stay at their current positions. The system dynamics are followed until a steady-state is established. The eventual number fraction of active particles $f_a$ is either zero or a steady positive value, depending on whether $\phi$ is below or above the critical volume fraction $\phi_c$. Unless otherwise specified, the amplitude of a kick is always randomly and uniformly chosen between 0 and $c\sqrt{\phi/N}$, where $N$ is the total number of particles and the constant $c$ is $1/2\sqrt{\pi}$ for disks and $1/4$ for noncircular particles. The system size used throughout this paper is 100,000 particles and ensemble averages of 10--100 configurations at the critical states are carried out for all of the reported results.
\begin{figure}[H]
\centering
\includegraphics[width=8cm, height=5cm]{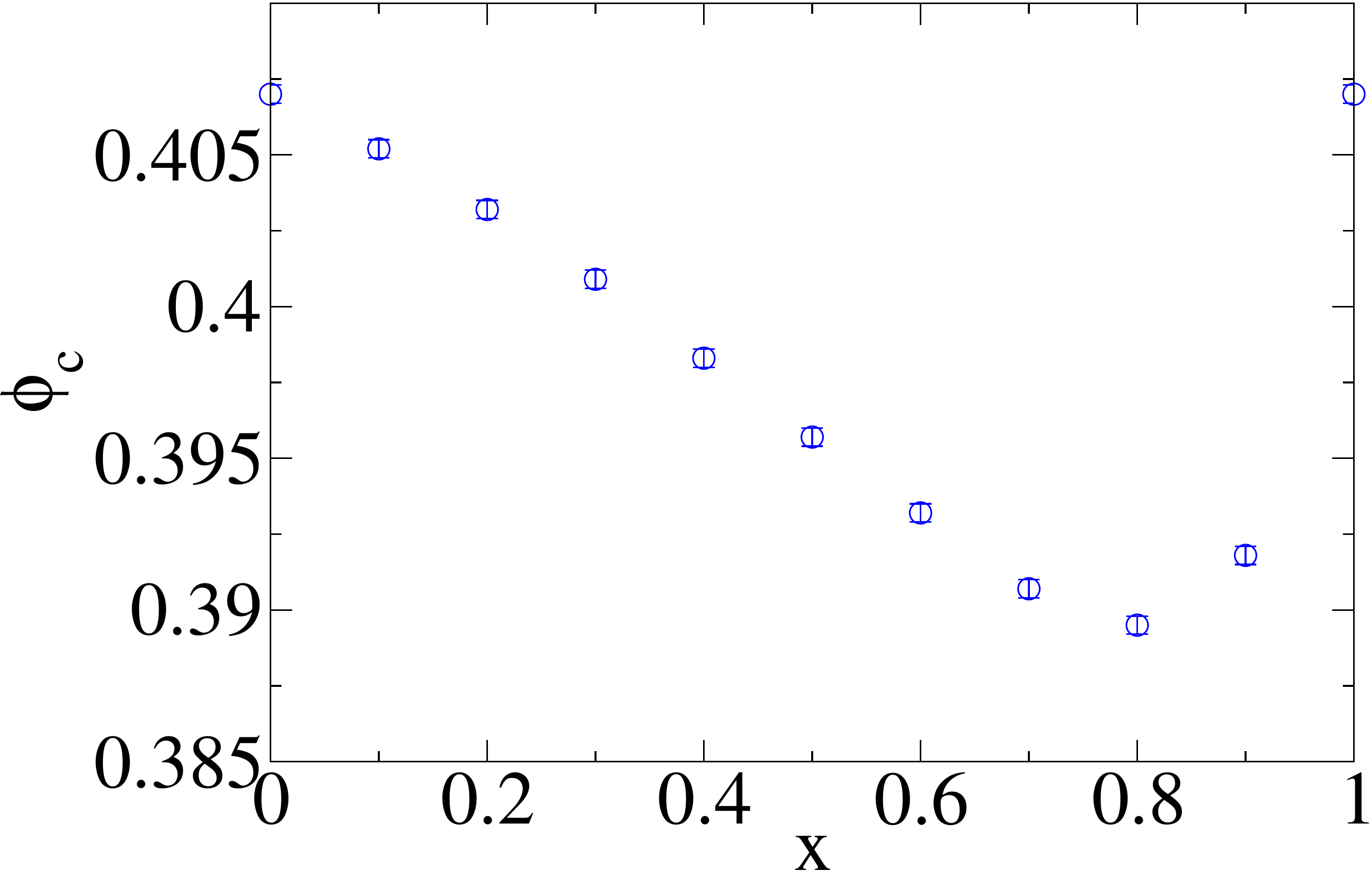}
\caption{Critical volume fraction $\phi_c$ as a function of the relative number fraction of small disks $x=N_S/(N_S+N_L)$ with fixed small to large particle size ratio $\gamma=0.5$.}
\label{fig: binary2}
\end{figure}

\section{Binary Mixtures of Disks}
\label{disks}
\begin{figure}[H]
\centering
\subfigure[\ Large]{
\includegraphics[width=8cm, height=5.5cm,clip=]{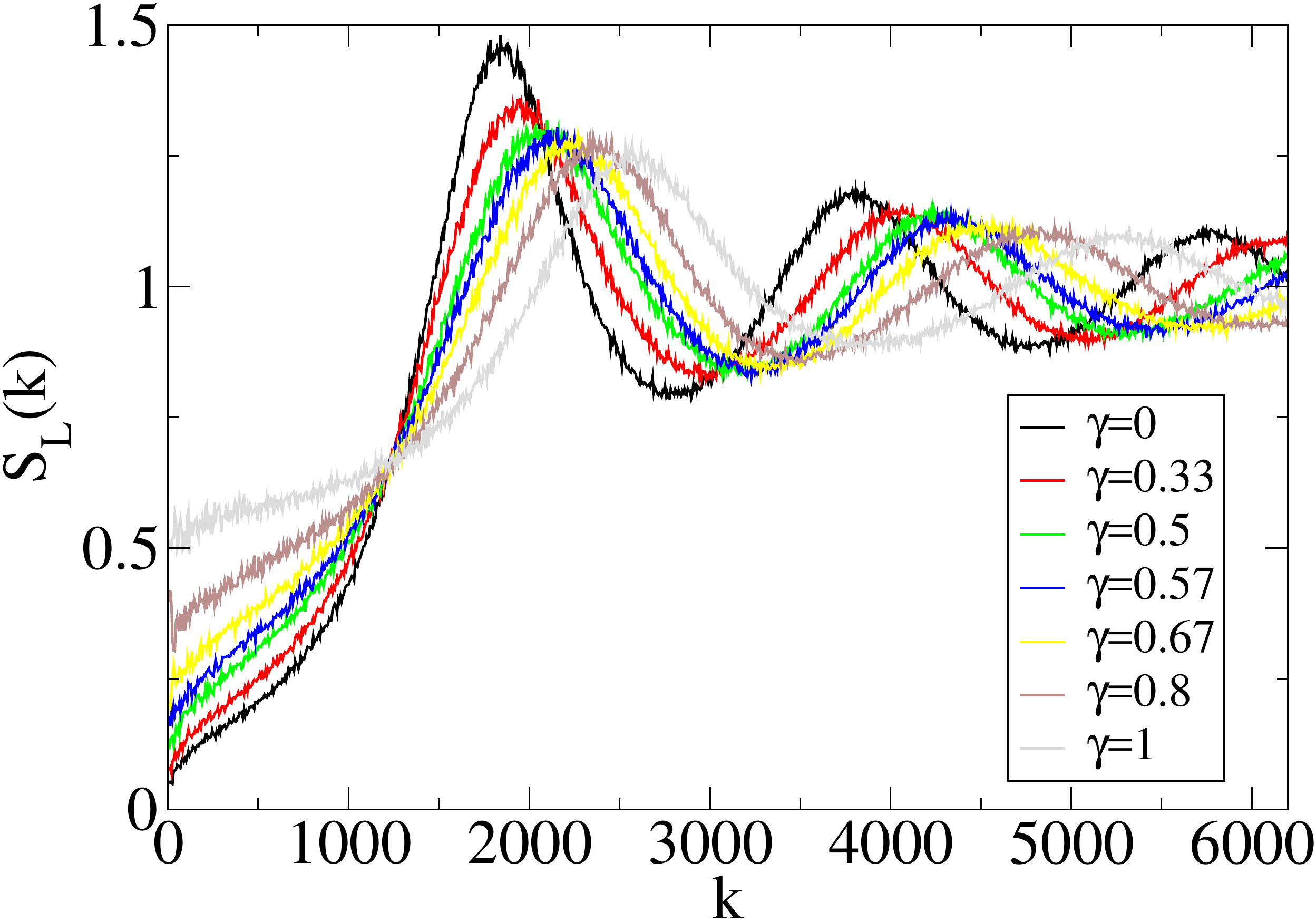}
}
\subfigure[\ Small]{
\includegraphics[width=8cm, height=5.5cm,clip=]{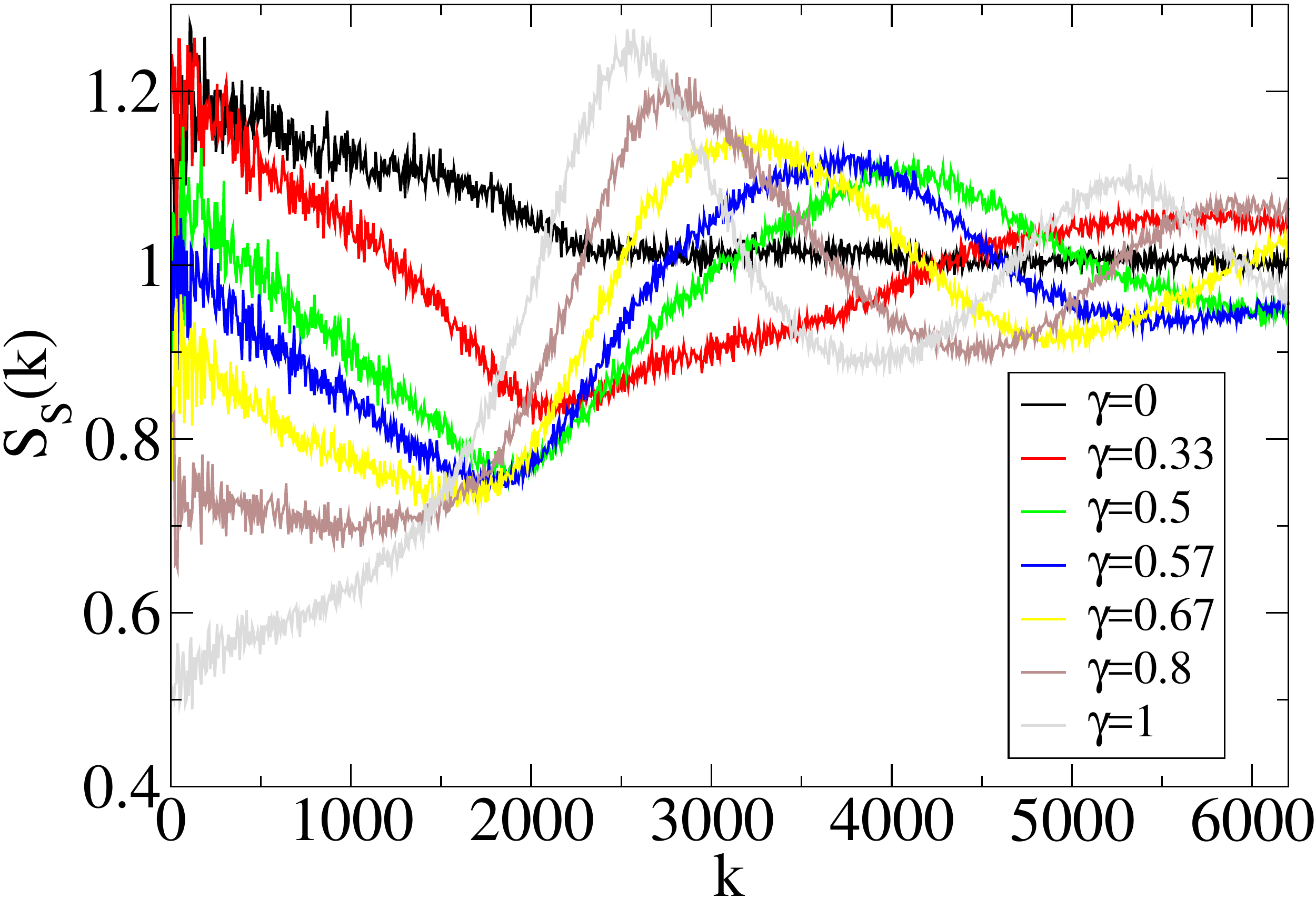}
}
\subfigure[\ Total]{
\includegraphics[width=8cm, height=5.5cm,clip=]{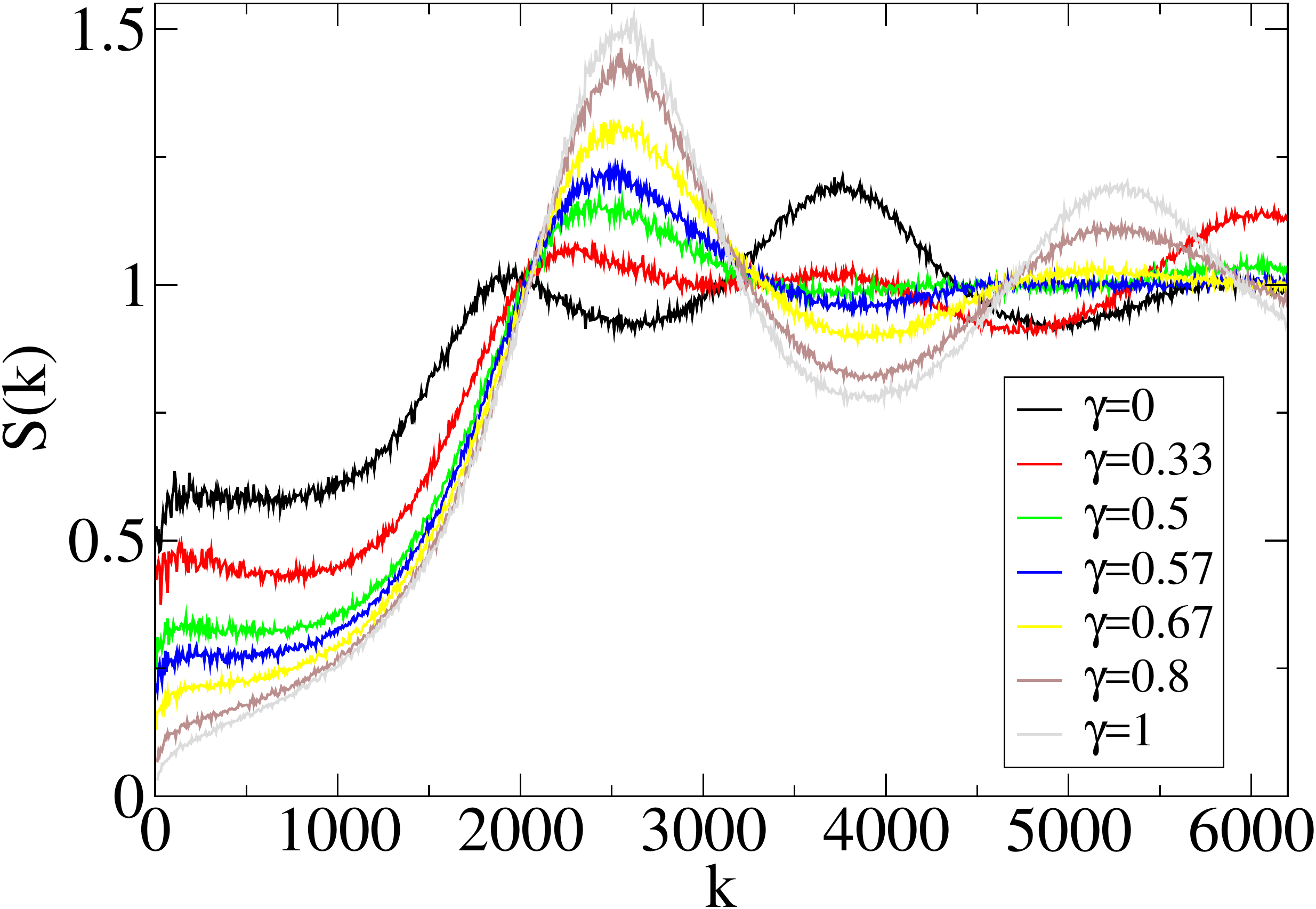}
}
\caption{Structure factors of the centers associated with  large particles $S_L(k)$, small particles $S_S(k)$ and the whole system $S(k)$ for different size ratios with fixed $x = 0.5$. Note that as the size discrepancy increases, large particles become more uniformly distributed while small particles become more disordered, but none of them are hyperuniform.}
\label{fig: nonhyper}
\end{figure}

We begin by considering two-dimensional models of binary mixtures of overlapping disks that under random organization dynamics lead to binary hard disks at the corresponding critical points. We let $\gamma=R_S/R_L$ denote the small-to-large particle size ratio and $x=N_S/(N_S+N_L)$ denote the relative number fraction of small disks. Since the parameter space is infinite, we restrict our study to two regimes: one in which we vary $\gamma$ at fixed $x=0.5$ and another in which we vary $x$ at fixed $\gamma=0.5$.\\  
\indent We first study the regime in which the small-disk fraction is fixed at $x=0.5$. We study a sequence of systems with $\gamma$ ranging from 0 to 1 and identify their critical absorbing states respectively. An example of a 2D binary disk system in an absorbing state is shown in Fig. \ref{fig: binary}(a). We find that the critical volume fraction $\phi_c$ varies nonmonotonically over the entire range of $\gamma$, as shown in Fig. \ref{fig: binary}(b). This also holds for systems with fixed $\gamma$ (see Fig. \ref{fig: binary2}), suggesting that the function $\phi_c(x,\gamma)$ is a smooth function bounded in a small interval. Note that in Fig. \ref{fig: binary}, as $\gamma$ decreases away from 1, $\phi_c$ decreases over a certain wide range of $\gamma$. However when $\gamma = 0$, the small particles become effectively points and the system behaves like a monodisperse system again. As $\gamma$ increases away from 0, $\phi_c$ decreases again such that there is a local minimum at around $\gamma=0.2$. A similar dip is observed at around $x=0.8$ in Fig. \ref{fig: binary2}.\\
\indent To ascertain the possible hyperuniformity exhibited by these critical absorbing states, we first compute the structure factors associated with the particle centroids for each species as well as for the whole system for these systems in Fig. \ref{fig: nonhyper}. As the size discrepancy increases ($\gamma$ decreases), the larger particles become more and more ordered, while the smaller particles goes in the opposite direction. The overall long-range density fluctuations increase as size discrepancy increases, as evidenced by a structure factor at the origin $S(k=0)$ that increases. Observe that there is always a small kink at small wave numbers.

\begin{figure}[H]
\centering
\subfigure[\ ]{
\includegraphics[width=8cm, height=5.5cm,clip=]{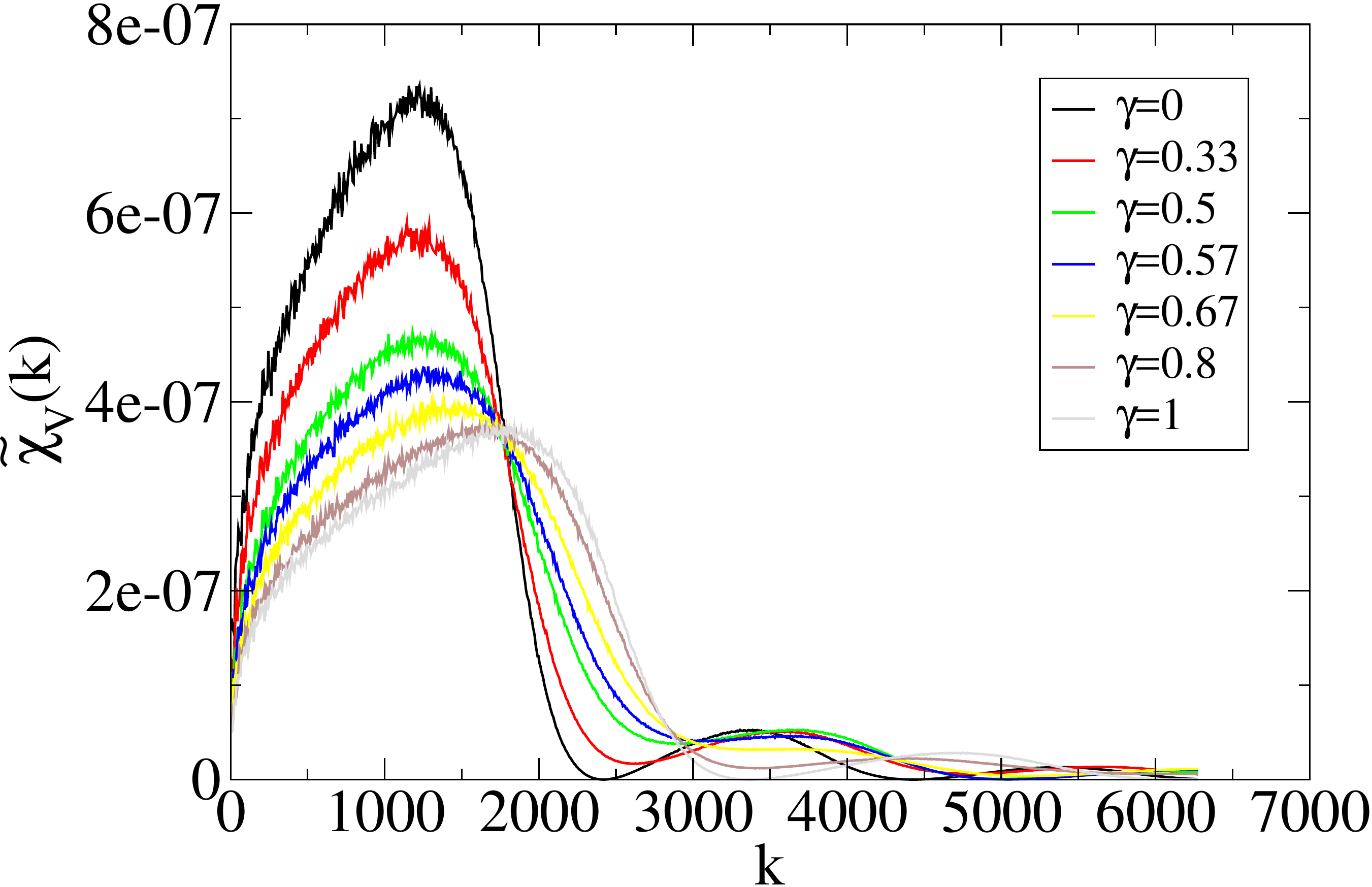}
}
\subfigure[\ ]{
\includegraphics[width=8cm, height=5.5cm,clip=]{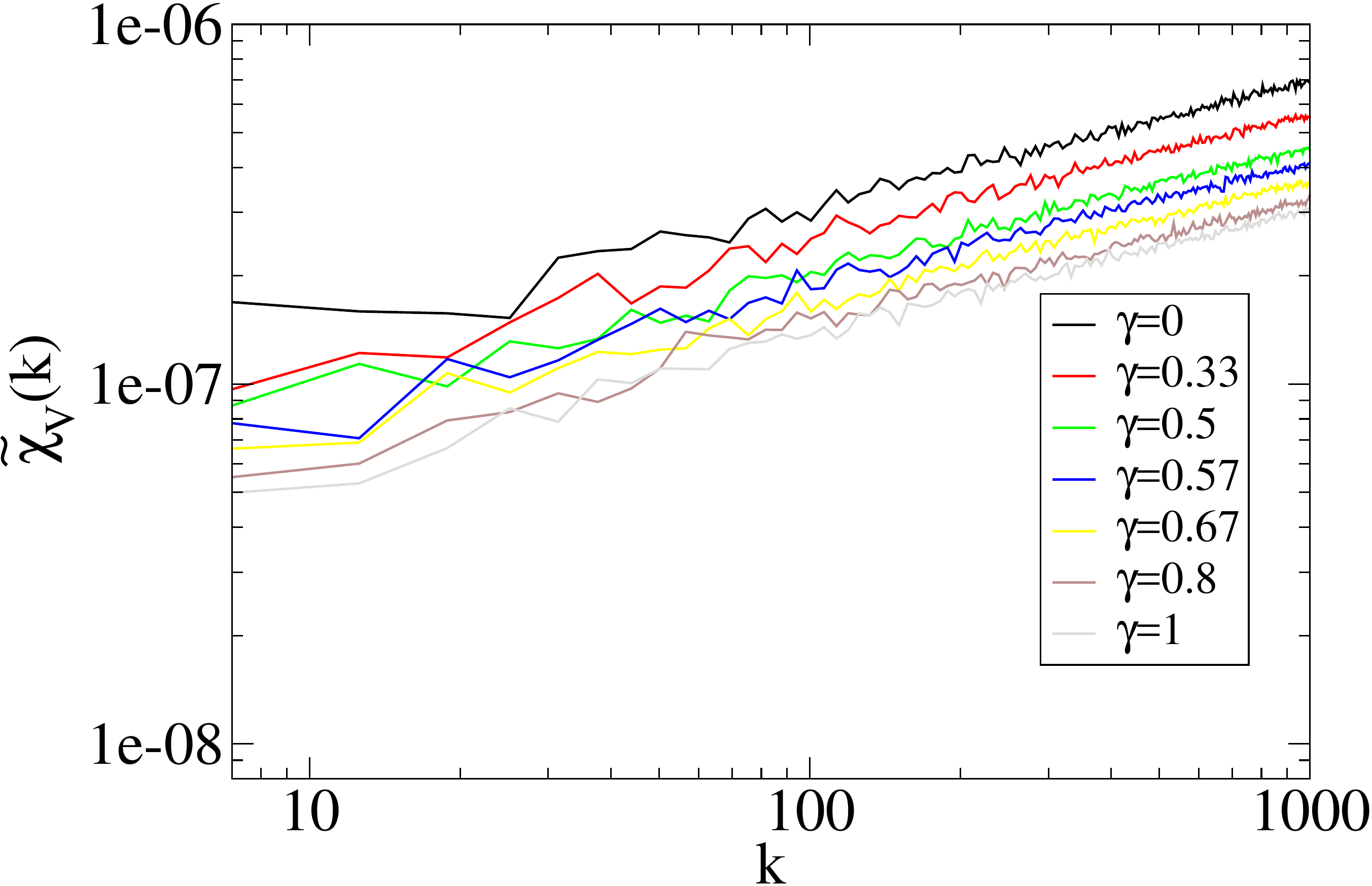}
}
\caption{(a) Spectral densities of binary-disk mixtures with different size ratios but fixed relative number fraction of small disks $x = 0.5$. (b) Small-$k$ behavior of the spectral densities presented in panel (a).}
\label{fig: chik}
\end{figure}
\begin{figure}[H]
\centering
\subfigure[\ Fixed $x=0.5$]{
\includegraphics[width=8cm, height=5.5cm,clip=]{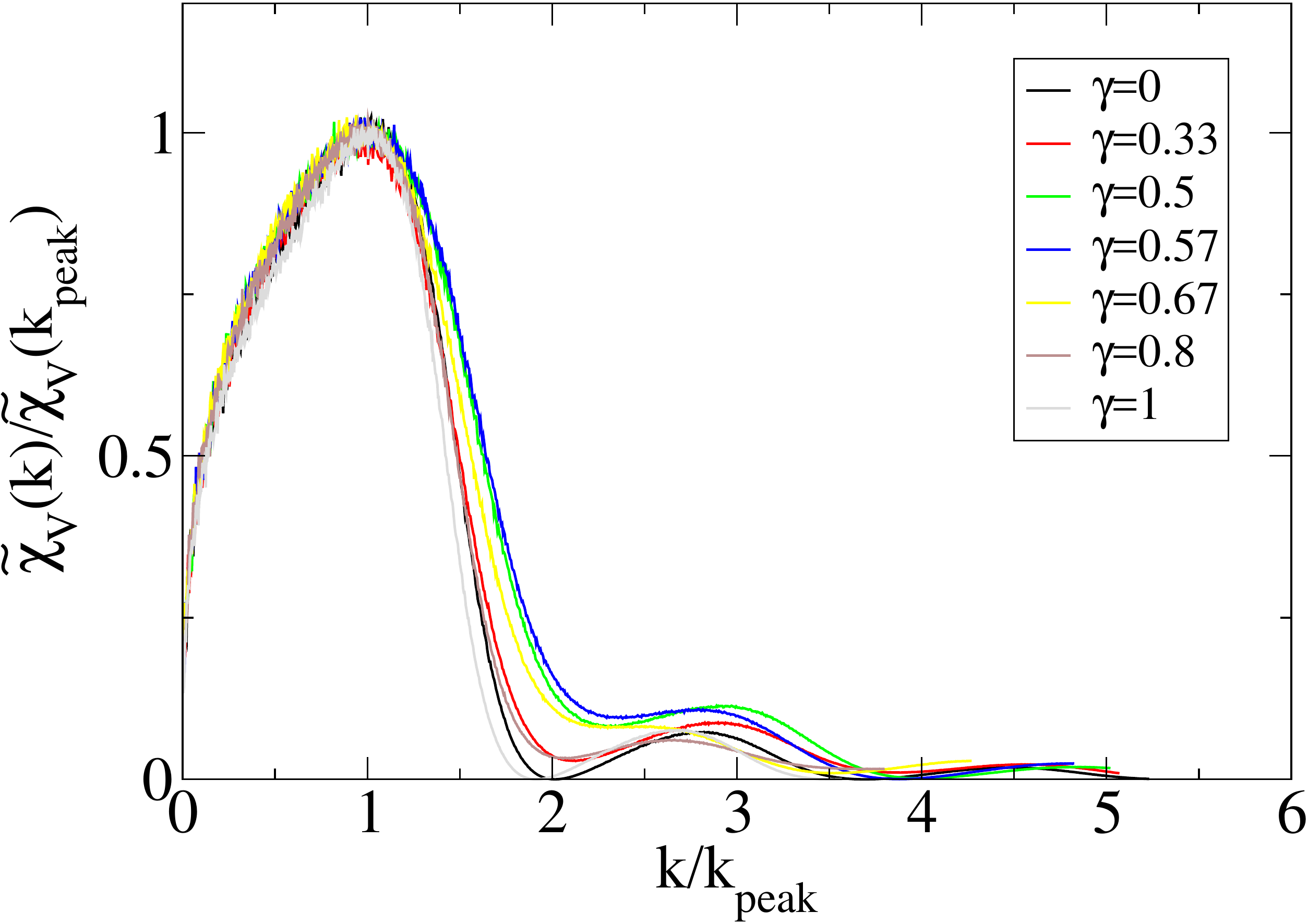}
}
\subfigure[\ Fixed $\gamma=0.5$]{
\includegraphics[width=8cm, height=5.5cm,clip=]{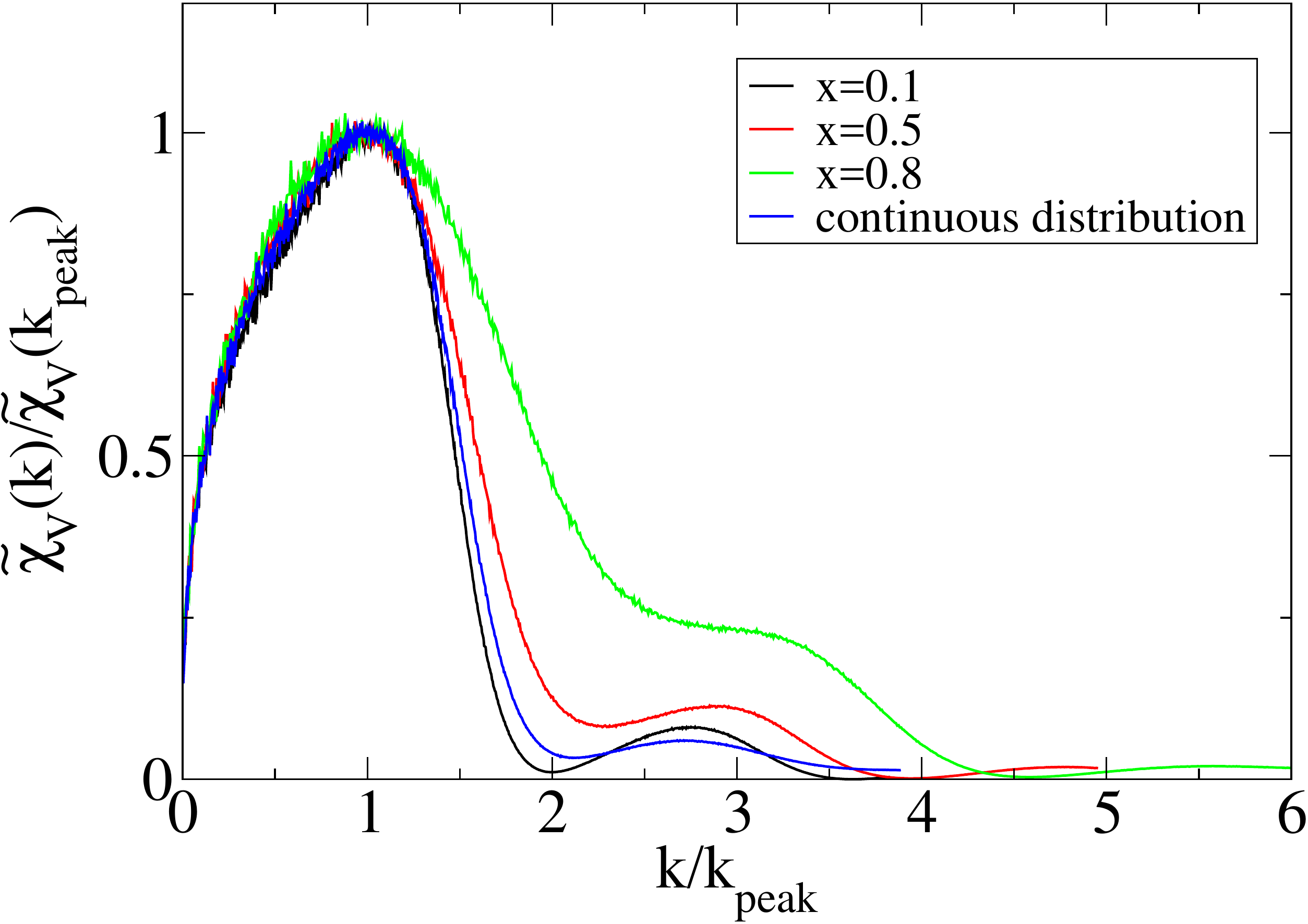}
}
\caption{Spectral densities rescaled by their first peaks [$k_{peak}$, $\tilde\chi_{_V}(k_{peak})$] for (a) binary-disk mixtures with fixed relative number fraction of small disks $x=0.5$ and (b) fixed small-to-large particle size ratio $\gamma=0.5$, as well as a mixture of disks with a continuous size distribution. Note that these curves approximately collapse onto a single curve in the small-$k$ region.}
\label{fig: rescaled}
\end{figure}

\indent Clearly, these binary systems are not hyperuniform if they are seen as point patterns determined by the particle centroids, as is the case for monodisperse disks. However, by applying the appropriate generalization of hyperuniformity to two-phase systems \cite{torquato2016hyperuniformity}, we find that the binary systems at criticality are indeed hyperuniform. We compute the spectral densities $\tilde\chi_{_V}(k)$ of the resulting two-phase systems by taking the space interior to particles as one phase and the space exterior as the other. The results are shown in Fig. \ref{fig: chik}. Notice that while the spectral densities for different compositions vary greatly, they all go to zero as $k \rightarrow 0$, meaning that they are all hyperuniform with respect to volume-fraction fluctuations. The scaling behavior of spectral densities, i.e., $\tilde\chi_{_V}(k) \sim k^{\alpha_{_V}}$ as $k \rightarrow 0$, is shown in the log-log plot of the small-$k$ region in Fig. \ref{fig: chik}(b). Interestingly, it further reveals that all of the spectral densities go to zero with the same scaling \footnote{For monodisperse spherical particles, we have $\tilde\chi_{_V}(k)=\rho(2\pi a/k)^dJ^2_{d/2}(ka)S(k)$ \cite{torquato2016hyperuniformity}, where $a$ is the radius and $J_{d/2}(ka)$ is the Bessel function. This guarantees that the spectral density scales in the same way as the structure factor as $k \rightarrow 0$, i.e., $\alpha_{_V}=\alpha_{_N}$. However, the relation between $\tilde\chi_{_V}(k)$ and partial structure factors becomes much more complex for polydisperse particles \cite{torquato2016hyperuniformity}, and one might not expect to see the same equality. Thus, it is a significantly more nontrivial result that polydisperse particles somehow arrange themselves in such a way that the spectral densities share the same scaling as the structure factor for monodisperse particles. Interestingly, this is what we see in Fig. \ref{fig: chik}(b) for different mixtures, despite the fact that their structure factors look quite different, as shown in Fig. \ref{fig: nonhyper}.}, implying that these systems have similar large scale structures despite their different compositions.\\ 
\indent Similarly, we find that for fixed small-to-large particle size ratio $\gamma=R_S/R_L=0.5$, although the spectral density profiles at the critical point change as the composition is varied, the spectral density profiles at the critical points have the same small-$k$ scaling. To better visualize this point, we show in Fig. \ref{fig: rescaled} the spectral densities rescaled by their first peaks for both fixed $x=0.5$ and $\gamma=0.5$. Note that these curves approximately collapse onto a single curve in the small-$k$ region, while for the region that $k>k_{peak}$ they vary greatly.

\indent We find the exponent $\alpha_{_V}=0.42\pm0.04$, which is consistent with previously reported values \cite{hexner2015hyperuniformity, tjhung2016criticality, weijs2015emergent}. We notice that the exponent $\alpha_{_V}$, given by the least square fit, is sensitive to the choice of fitting intervals due to the fluctuations in the small-$k$ region. In order to obtain a reliable value, we ran over 100 samples of the monodisperse system consists of 100,000 particles, and averaged over the results of 87 samples that finally reached absorbing states. We find that as the fitting interval shrinks from the first 50 $k$-points to the first 5 $k$-points, the best fitted exponent gradually increases from 0.38 to 0.45; however, the corresponding uncertainty increases from $\pm0.005$ to $\pm0.1$. To balance the bias induced by the large intervals and the fluctuations in the small intervals, we report the result $0.42\pm0.04$, which lies in the middle of the two extremes.\\ 
\indent The hyperuniformity of the resulting two-phase media, as quantified by the spectral density, along with the nonhyperuniformity of particle centroids, implies that the rearrangement of particles to suppress their volume-fraction (``mass") fluctuations rather than the number fluctuations is at the heart of random organization dynamics. Since the former implies the latter in the monodisperse case, this discovery could not have been made if we had treated number fluctuations associated with the centroids of the particles, as was done in previous studies of monodisperse spheres (circles).  

\begin{figure*}[]
\centering
\subfigure[\ Disks]{
\includegraphics[width=5.5cm, height=5.5cm]{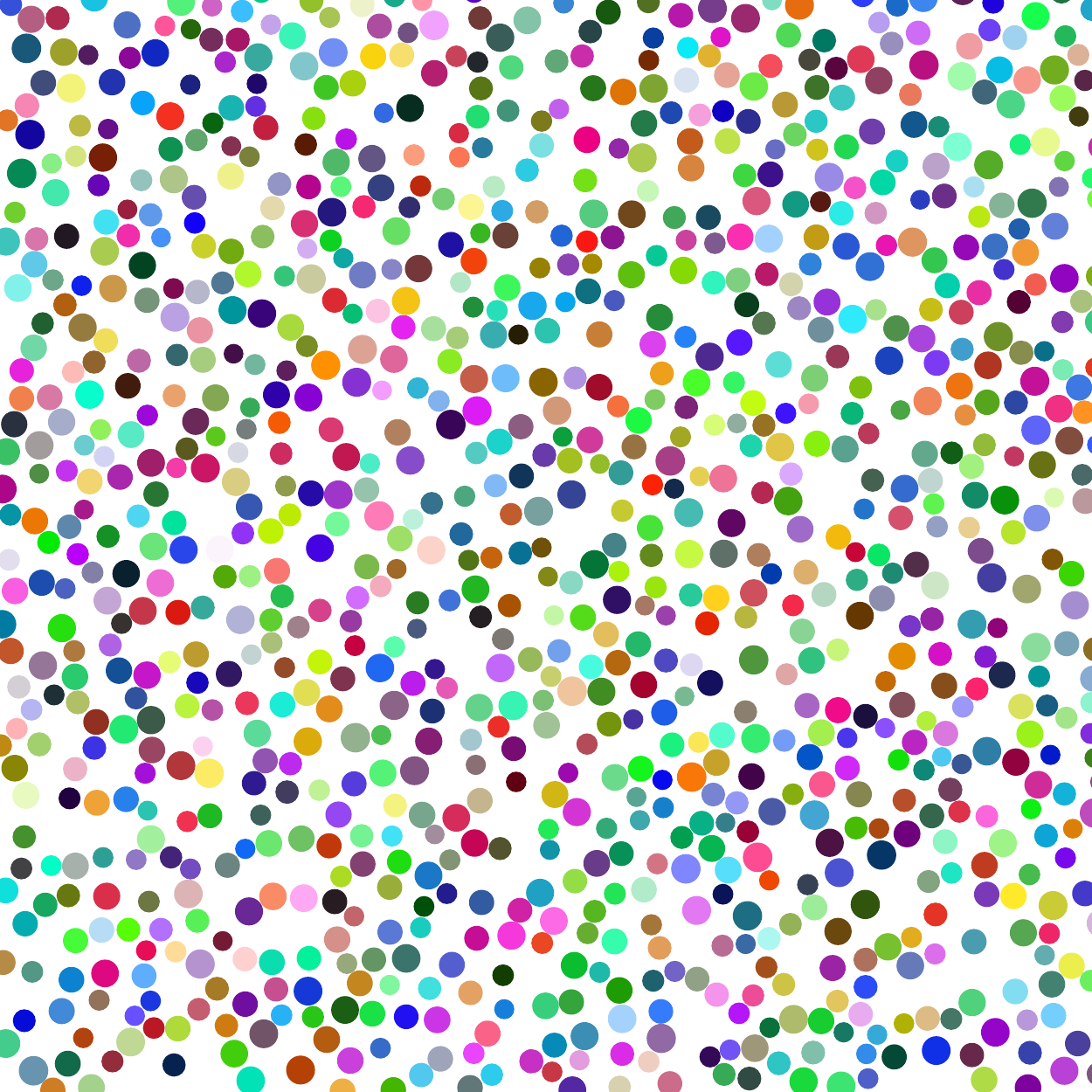}
}
\subfigure[\ Squares]{
\includegraphics[width=5.5cm, height=5.5cm]{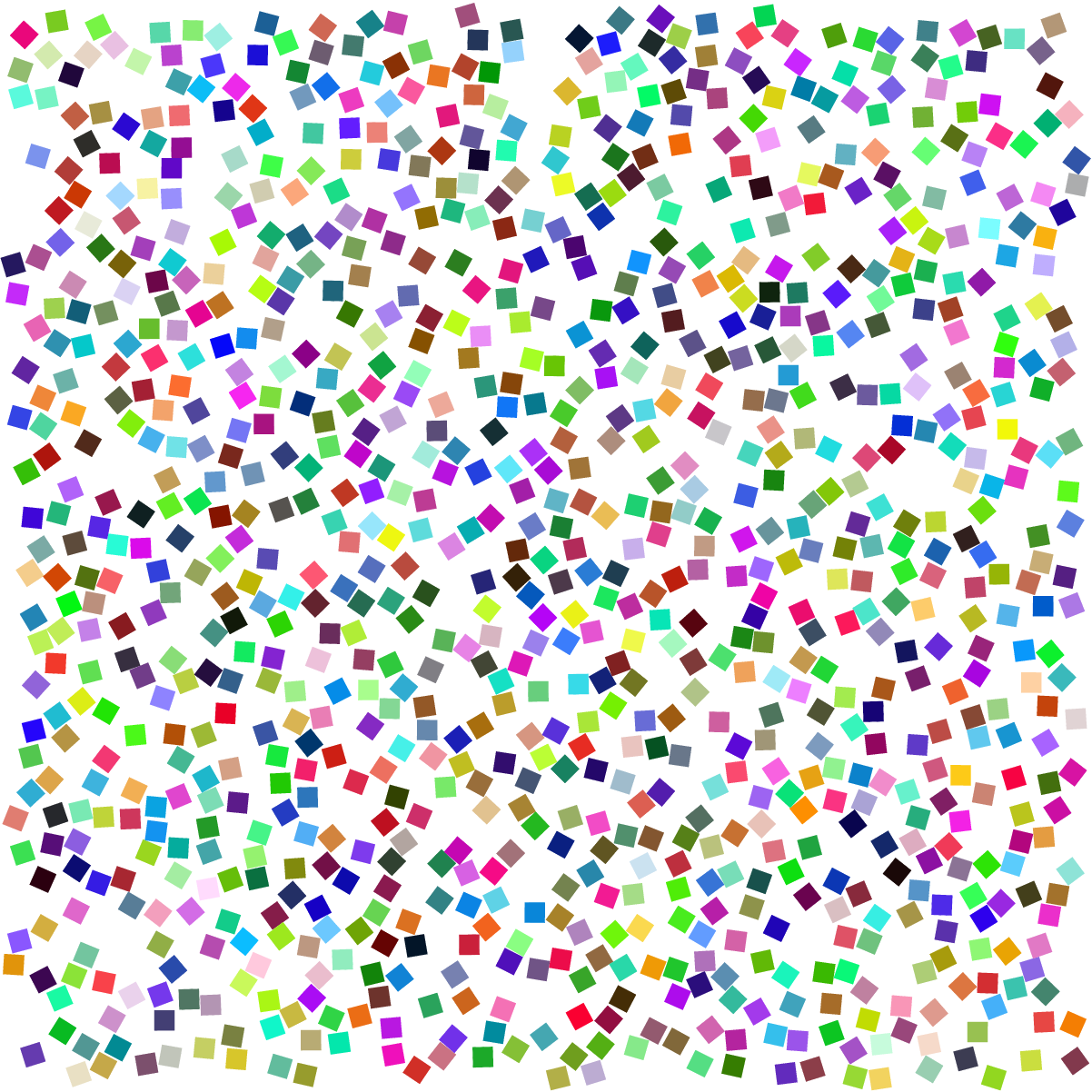}
}
\subfigure[\ Needles]{
\includegraphics[width=5.5cm, height=5.5cm]{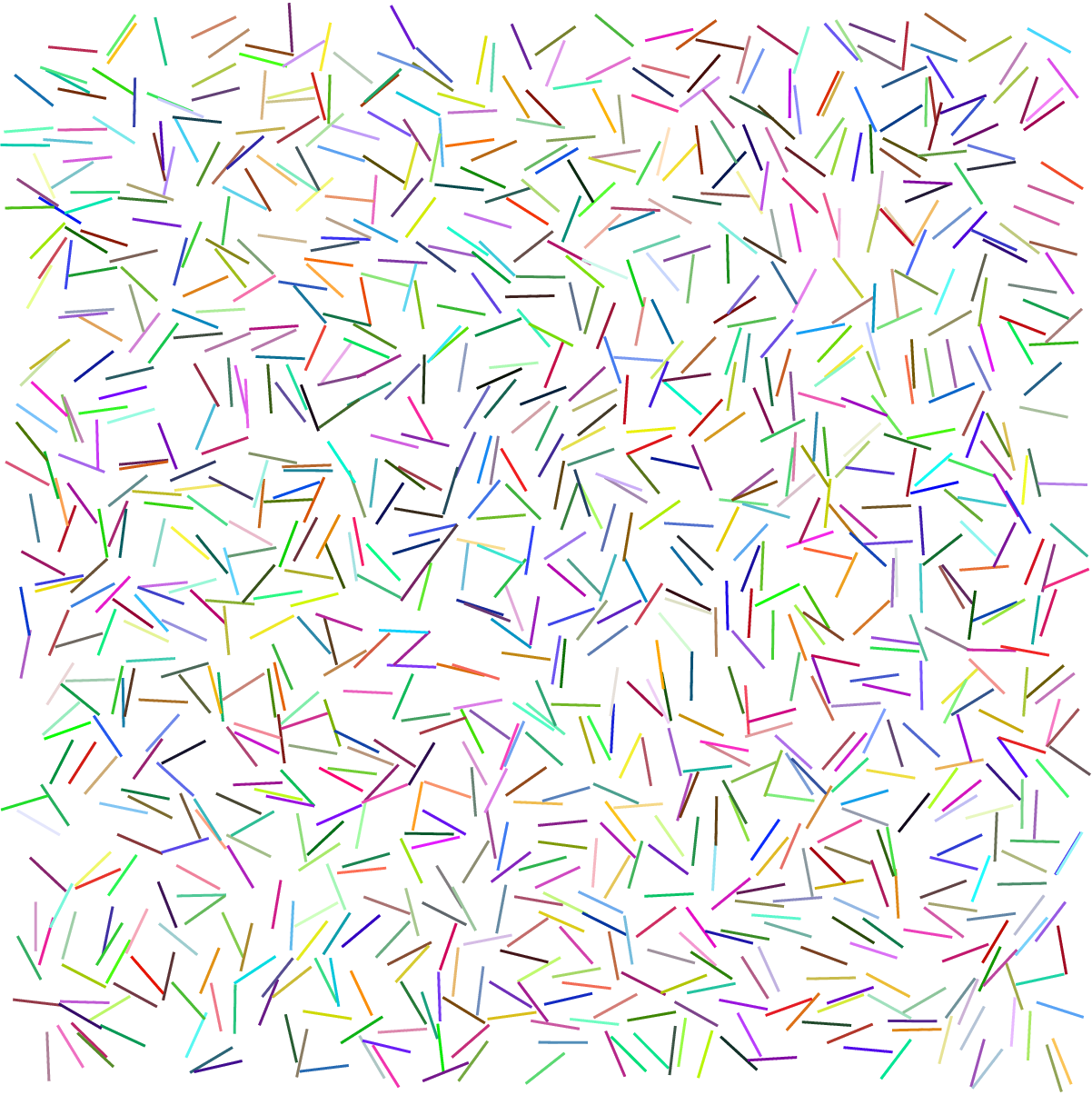}
}
\subfigure[\ Critical reduced density]{
\includegraphics[width=5.2cm, height=3.8cm]{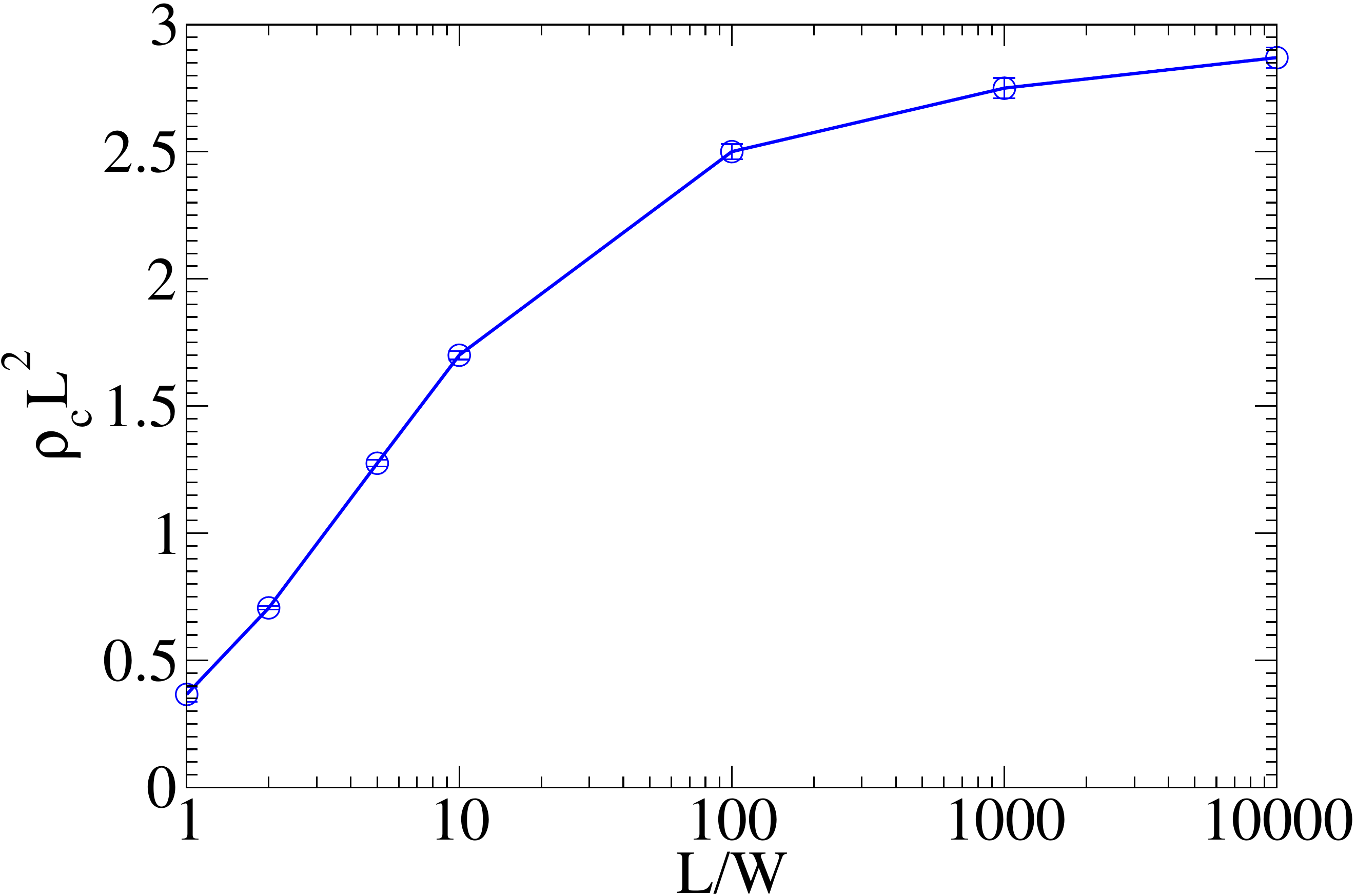}
}
\subfigure[\ Orientational correlation function]{
\includegraphics[width=5.2cm, height=3.8cm]{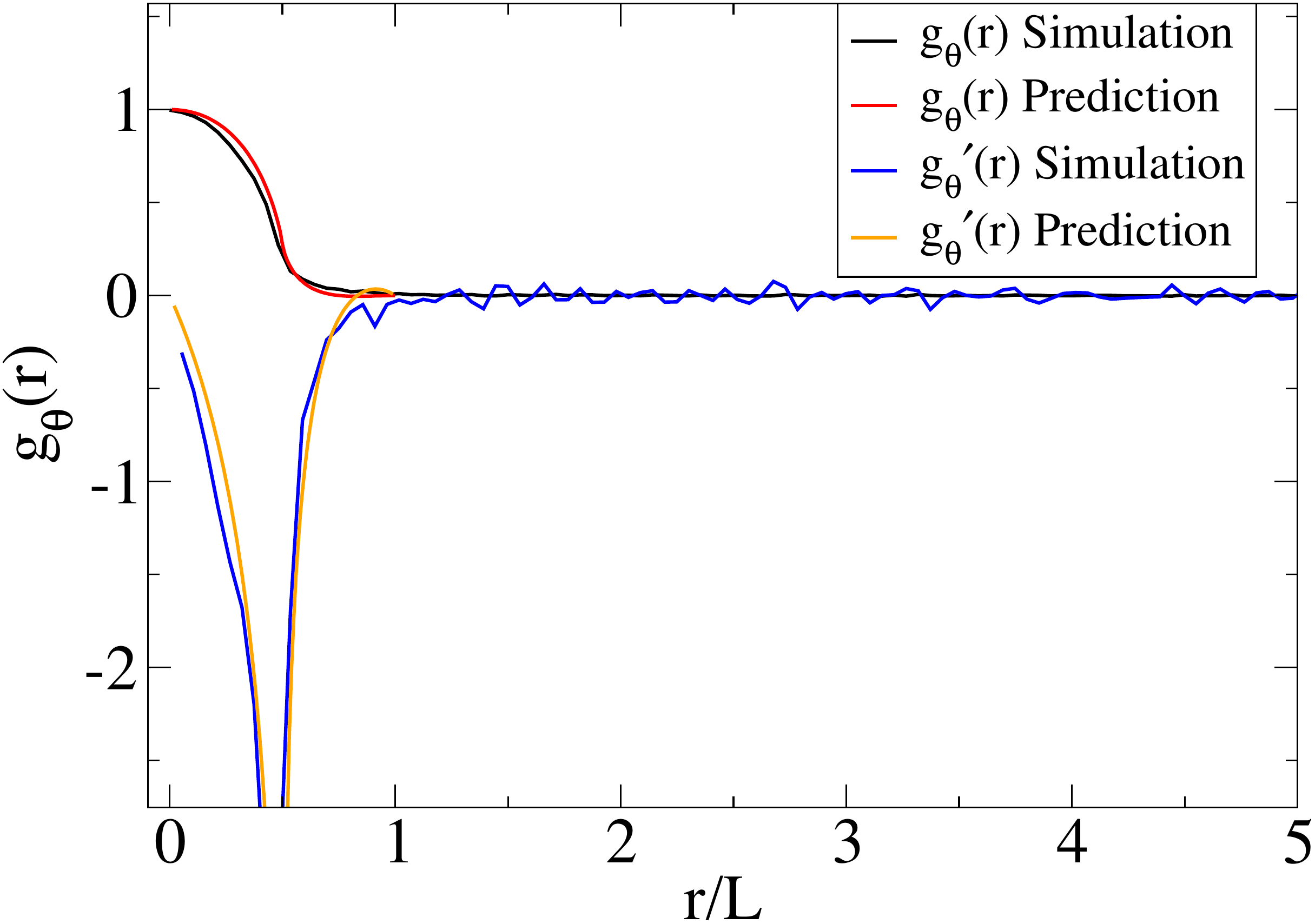}
}
\subfigure[\ Spectral densities]{
\includegraphics[width=6cm, height=4.5cm]{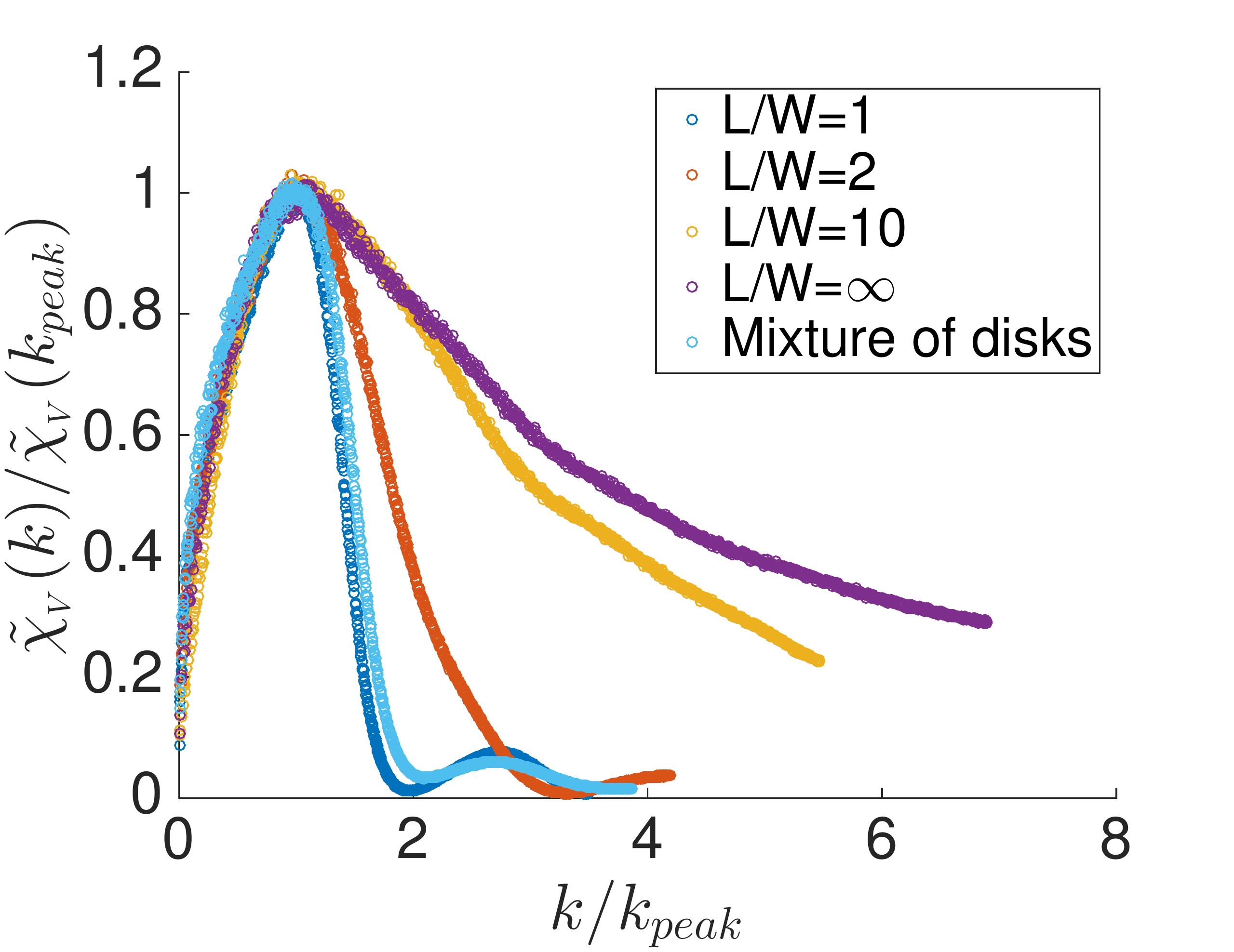}
}
\caption{(a) A representative image of a configuration of a 2D-disk system with continuous particle-size distribution in its critical absorbing state. (b) A representative image of a 2D-square system in its critical absorbing state. (c) A representative image of a configuration of a 2D-hard-needle system in its critical absorbing state. (d) The critical reduced density $\rho_cL^2$ as a function of the aspect ratio $L/W$ for rectangles, $L/W=1$ corresponds to the image in panel (b) and $L/W=\infty$ corresponds to the image in panel (c). (e) Simulated and predicted [Eq. (\ref{theory})] orientational correlation function $g_{\theta}(r)$ as well as their corresponding derivatives for hard-needle systems at low densities. Notice that the second derivative has a discontinuity at $r=L/2$ for both ones. (f) Comparison of spectral densities rescaled by their first peaks [$k_{peak}$, $\tilde\chi_{_V}(k_{peak})$] for corresponding configurations in panel (a) (mixture of disks), (b) ($L/W=1$), and (c) ($L/W=\infty$) near the critical point.}
\label{fig: nonspherical}
\end{figure*}
To further generalize our findings, we investigate disks with a continuous size distribution (with radius chosen from a uniform distribution $U(R,1.5R)$); see Fig. \ref{fig: nonspherical}(a) for a representative image. We find again that the two-phase medium at the critical point is hyperuniform, as shown in Fig. \ref{fig: nonspherical}(f).\\

\section{Nonspherical/Noncircular Particle Shapes}
\label{non}

\begin{figure}[H]
\centering
\includegraphics[width=8cm, height=6cm]{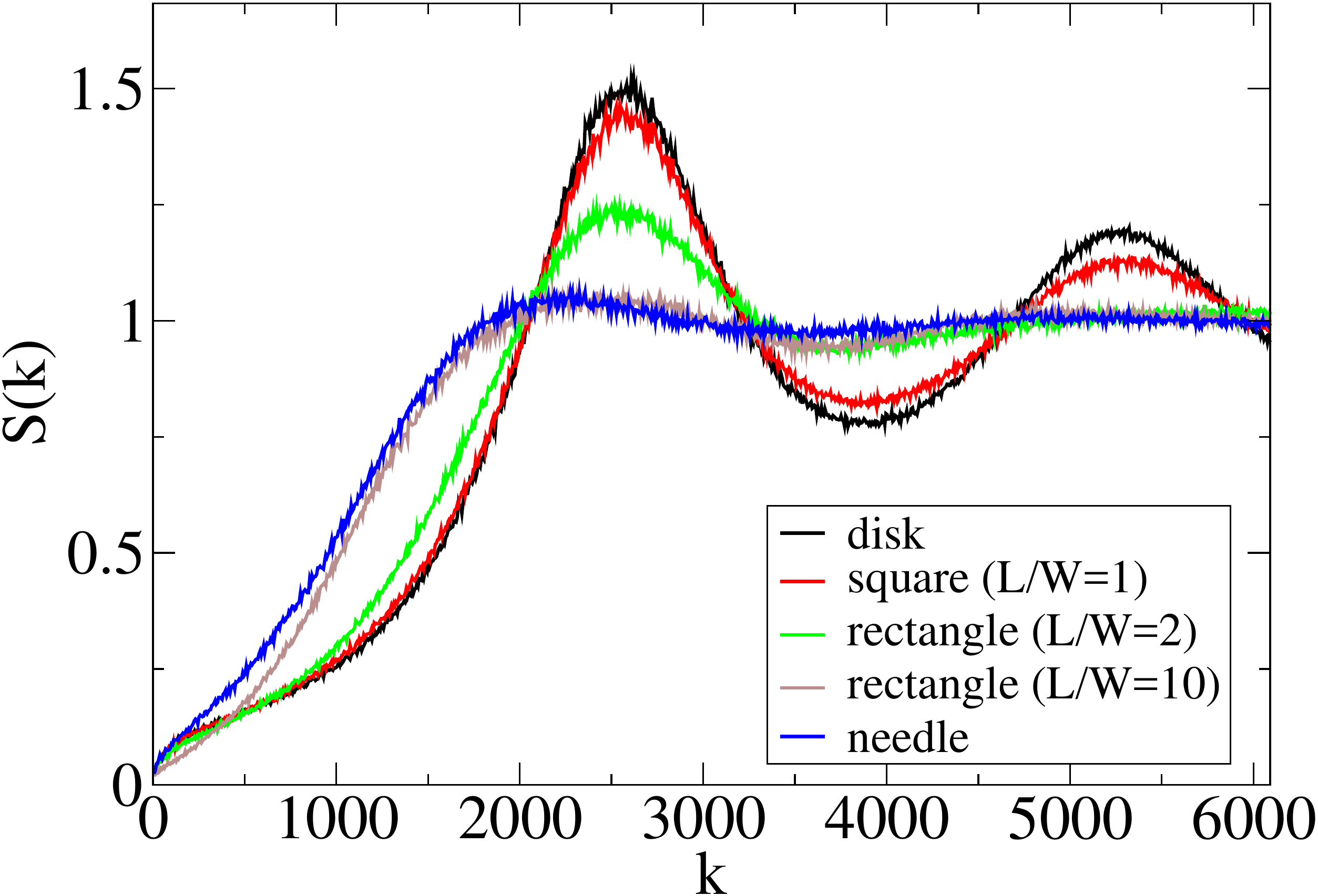}
\caption{Comparison of structure factors for configurations of monodisperse particles with different shapes near their corresponding critical points under the random organization dynamics.}
\label{fig: nonstructure}
\end{figure}

Our findings thus far have revealed that the random organization mechanism provides a robust means of generating hyperuniform systems that consist of a variety of mixtures of spherical (circular) particles. This naturally leads us to consider a wider class of particle shapes. Specifically, we extend this model to noncircular particles \cite{onsager1949effects}, namely hard rectangles of certain aspect ratios $L/W$ (where $L$ is the length and $W$ is the width of the rectangle), including the hard-needle limit. This allows us to study the effect of rotational degree of freedom. For every random kick, we also need to include a random rotation $\delta \theta$ of the active particle. Here we chose $\delta \theta$ uniformly from $[-\theta_0,\theta_0]$. We did not find fundamental differences between different choices of $\theta_0$ except for a shift of the critical point, and thus we use $\theta_0=\pi/2$ for the ensuing discussion.\\
\indent Representative images of squares ($L/W=1$) and needles ($L/W=\infty$) in their corresponding critical absorbing states are shown in Figs. \ref{fig: nonspherical}(b) and (c). For rectangles, we find that the critical packing fraction $\phi_c$ decreases monotonically as the aspect ratio $L/W$ increases. This is not surprising, since in the limit $L/W \rightarrow \infty$, the problem is reduced to the random organization of hard needles. Since the packing fraction is diminishing to zero in this limit, a more proper quantity to focus on is $\rho_c L^2$, which we refer to as a critical reduced density. We plot the critical reduced density as a function of $L/W$ in Fig. \ref{fig: nonspherical}(d), and one can see that it converges as $L/W \rightarrow \infty$. \\ 
\indent Unlike the quasi-long-range correlation found in the XY model \cite{kosterlitz1973ordering} and two-dimensional hard rods at equilibrium \cite{bates2000phase}, the orientational correlation function
\begin{equation} \label{eq: orient}
g_{\theta}(r)=\left\langle \cos(2(\theta(0)-\theta(r)))\right\rangle
\end{equation}

\noindent for hard needles decays to zero almost immediately beyond $r=L$ (see Fig. \ref{fig: nonspherical}(e)). This definitively demonstrates that only short-range orientational order is present in these noncircular particle systems. Note that the critical reduced density is well below the isotropic-nematic transition density \cite{frenkel1985evidence, kayser1978bifurcation}. Indeed, we did not observe any nematic order in these systems.\\ 
\indent We have also theoretically determined the orientational correlation function based on hard needles in equilibrium in the dilute limit. We assume the needle has unit length and the distance between the centroids of two needles is $r$. Let 
\begin{align}
&\theta_1(x;r)=\sin ^{-1}\left(\frac{r \sin (x)}{\sqrt{r^2-r \cos (x)+0.25}}\right),\\
&\theta_2(x;r)=\sin ^{-1}(2 r \sin (x)).
\end{align}
By averaging over all of the feasible configurations of two hard needles, the orientational correlation function $g_{\theta}(r)=\left\langle \cos(2(\theta(0)-\theta(r)))\right\rangle$ can be written as 
\begin{widetext}
\begin{equation} \label{theory}
g_{\theta}(r)= \left \{ \begin{aligned} 
&\frac{\int_{0}^{\cos ^{-1}(r)}\frac{\sin \left(2 \theta_1(x;r)\right)+\sin \left(2 \theta_2(x;r)\right)}{\theta_1(x;r)+\theta_2(x;r)}dx+\int_{\cos ^{-1}(r)}^{\frac{\pi }{2}}\frac{\sin \left(2 \theta_2(x;r)\right)}{\theta_2(x;r)}dx}{\pi }, &0<r\leq\frac{1}{2},\\ 
&\frac{\int_{0}^{\cos ^{-1}\left(\frac{1}{2 r}\right)}\frac{\sin \left(2 \theta_2(x;r)\right)-\sin \left(2 \theta_1(x;r)\right)}{\pi-\theta_1(x;r)+\theta_2(x;r)}dx+\int_{\cos ^{-1}\left(\frac{1}{2 r}\right)}^{\cos ^{-1}(r)}\frac{\sin \left(2 \theta_1(x;r)\right)+\sin \left(2\theta_2(x;r)\right)}{\theta_1(x;r)+\theta_2(x;r)}dx+\int_{\cos ^{-1}(r)}^{\sin ^{-1}\left(\frac{1}{2 r}\right)}\frac{\sin \left(2 \theta_2(x;r)\right)}{\theta_2(x;r)}dx}{\pi }, &\frac{1}{2}<r\leq \frac{\sqrt{2}}{2},\\
&\frac{\int_{0}^{\cos ^{-1}(r)}\frac{\sin \left(2 \theta_2(x;r)\right)-\sin \left(2 \theta_1(x;r)\right)}{\pi-\theta_1(x;r)+\theta_2(x;r)}dx}{\pi }, &\frac{\sqrt{2}}{2}<r\leq1,\\ 
&0 &r>1.\end{aligned}\right.    
\end{equation}
\end{widetext}
We find that the predictions of our formula agrees well with our simulations, as shown in Fig. \ref{fig: nonspherical}(e).\\
\indent The spectral densities (see Fig. \ref{fig: nonspherical}(f)) show that for these noncircular particles, hyperuniformity is still preserved. For hard needles, it is crucial to employ the spectral density for interface to obtain the result \cite{torquato2016hyperuniformity, ma2018precise}. Interestingly, as the aspect ratio $L/W$ increases, one can clearly see the destruction of short-range order from the increasingly diffusive tails in the spectral densities. This can also be seen from the damped oscillations in the structure factors of the centroids for the particles (see Fig. \ref{fig: nonstructure}). \footnote{The structure factor of the centroids for squares is very similar to that of disks in the sense that the positions of the peaks are unchanged but with oscillations that are somewhat damped. We believe this is a common feature for random organization models of all regular polygons.} The reason is that less symmetric particles would broaden the distribution of short-range particle-particle spacing. However, as one can see from the Fig. \ref{fig: nonspherical}(f), despite the fact that the spectral density profiles appear to be quite different from one another, they all exhibit the same small-$k$ behavior. \footnote{For $L/W\geq10$, it is computationally very challenging to precisely identify the critical point. A larger system is required for this large aspect ratio, which makes it harder to ascertain the small-$k$ behavior of the spectral density. Indeed, simulations for hard needles are actually carried out using a different choice of the largest random kick, namely, $L/4$ rather than the choice we used for other rectangular shapes, which would yield a vanishing random kick due to the vanishing packing fraction.}

\section{Discussion} 
\label{discuss}
\begin{figure}[ht]
\centering
\subfigure[]{
\includegraphics[width=7cm, height=5cm,clip=]{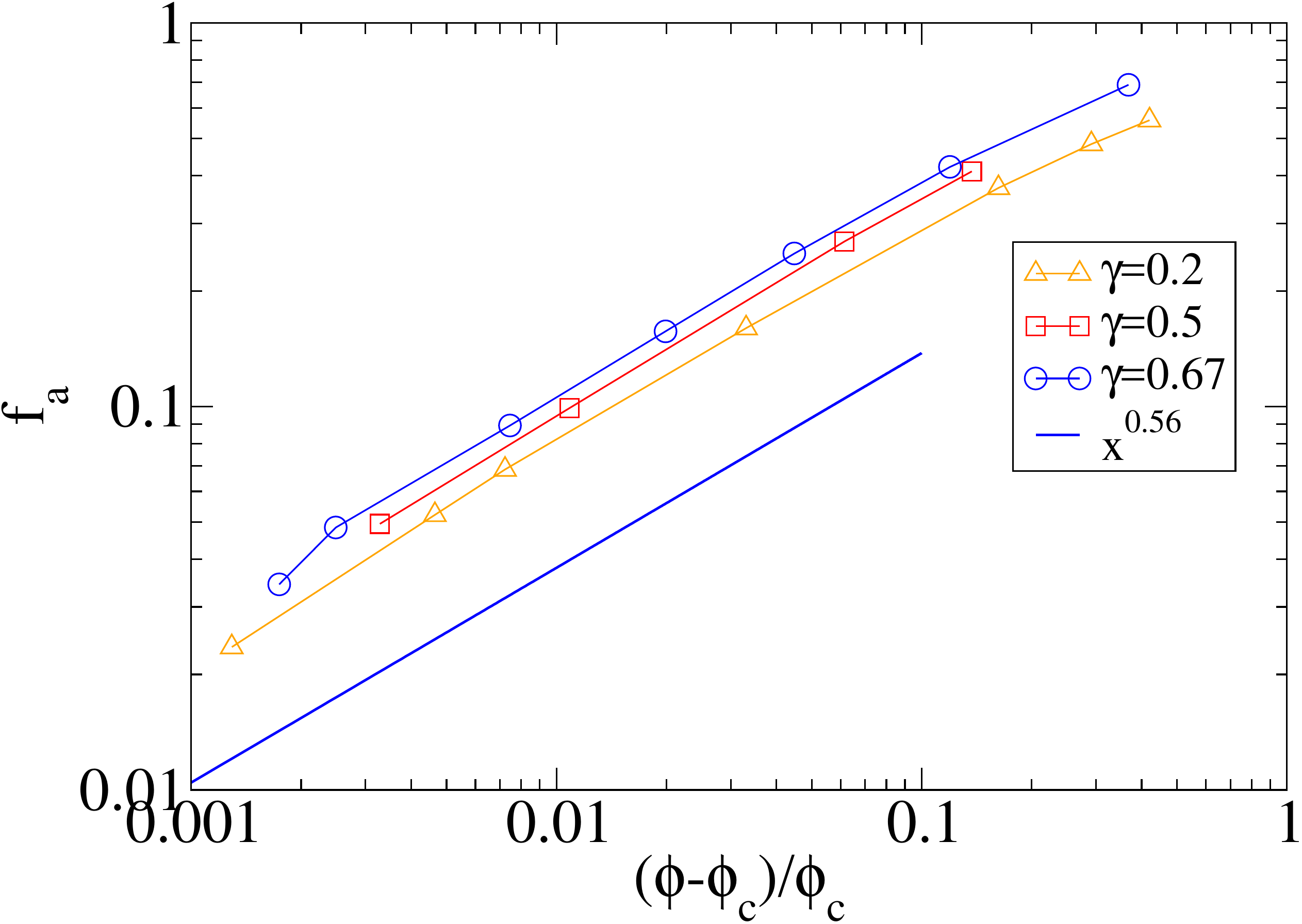}
}
\subfigure[]{
\includegraphics[width=7cm, height=5cm,clip=]{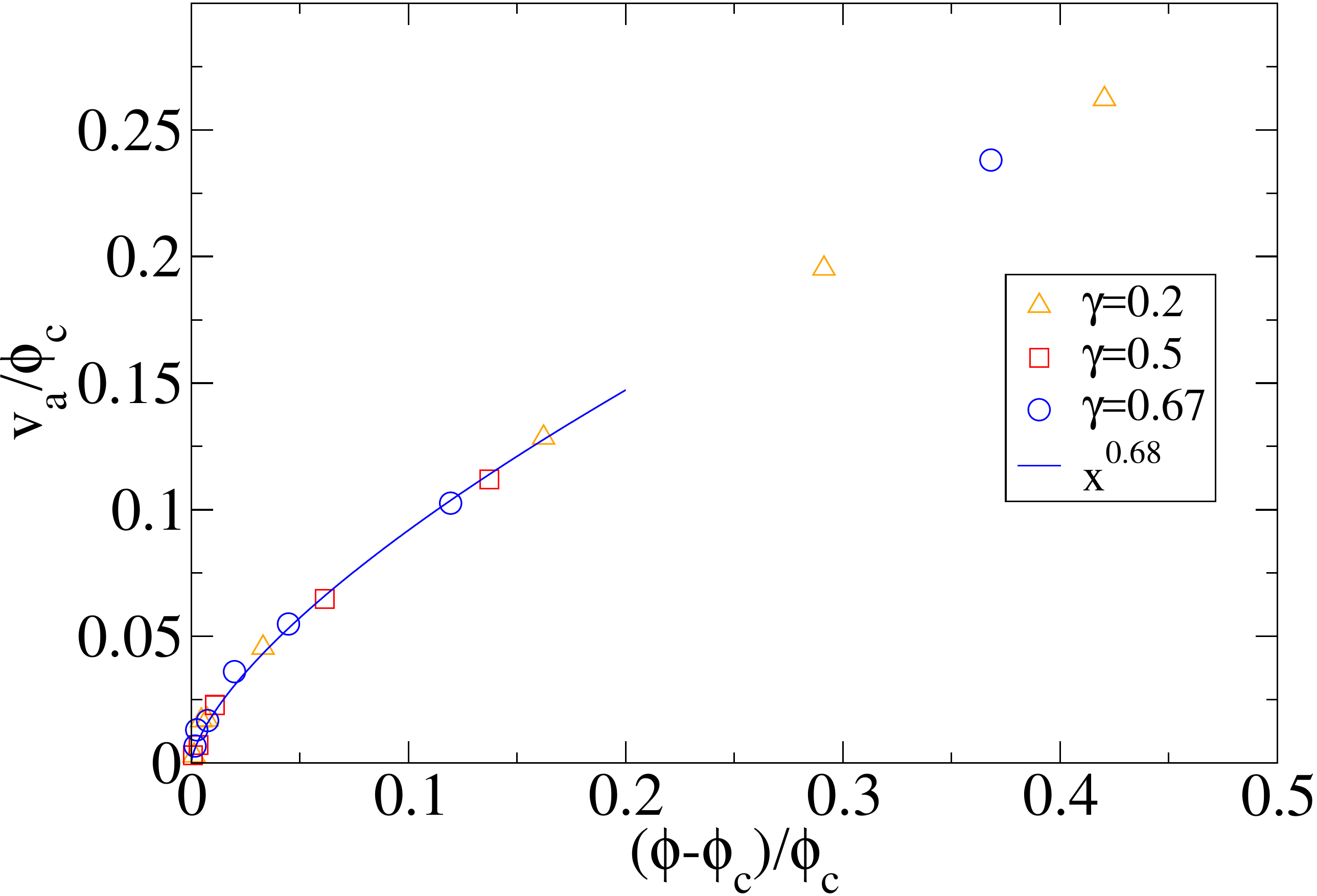}
}
\caption{(a) The number fraction of active particles $f_a$ and (b) the rescaled active volume fraction $v_a$ as functions of the scaled volume fractions for different small to large particle size ratios $\gamma$ with $x=0.5$. }
\label{fig: exponent}
\end{figure}

The results of our investigation expand our understanding of random organization as a model for absorbing phase transitions of continuous media by going beyond the previously studied monodisperse systems. Our analysis reveals that the redistribution of the ``mass" of the particles rather than the particle centroids is central to this dynamical process.\\ 
\indent The more general treatment of the systems as two-phase media leads us to the proposition that the ``active volume fraction" $v_a$ (i.e., the volume fraction of the total common volume between overlapping particles) is the appropriate general quantity to examine rather than the number fraction of active particles $f_a$. The use of $v_a$ may help to resolve the undetermined universality class of the absorbing phase transition exhibited by the random organization model \cite{corte2008random, hexner2015hyperuniformity, tjhung2016criticality}. Here we show some preliminary results that support this possibility. We have computed the final $f_a$ as a function of $(\phi-\phi_c)/\phi_c$ for a few systems in Fig. \ref{fig: exponent}(a). We find that the critical exponent $\beta \simeq 0.56 \pm 0.02$ ($f_a \propto (\phi-\phi_c)^\beta$), which is close to the value for directed percolation in 2+1 dimensions \cite{lubeck2004universal}, as suggested by previous work \cite{tjhung2016criticality}. However, although for different mixtures, the values of $f_a$ have approximately the same scaling, those curves never collapse onto a single one. Interestingly, if the number fraction of active particles is replaced by active volume fraction $v_a$, then the corresponding curves collapse onto a single one, as one can see in Fig. \ref{fig: exponent}(b). This implies that the active volume fraction is a more appropriate quantity to employ and investigate. Moreover, by using the active volume fraction, we find a different exponent $\beta' \simeq 0.68 \pm 0.08$, which is closer to the value of the universality class of conserved directed percolation \cite{lubeck2004universal}. This agreement and the fact that the particle number and mass are conserved suggest that conserved directed percolation may be the correct universality class to which random organization models belong. This remains an outstanding question for future research.
\bigskip

\indent Our work also demonstrates that random organization provides a robust and versatile means of generating a wide class of disordered hyperuniform two-phase media. This includes a broad spectrum of mixtures of particles (discrete and continuous) of circular or noncircular shape. Practically, our findings imply that hyperuniform materials can be made in the laboratory without the need to use monodisperse particles, which eases the preparation process. Moreover, different structures can be made in the laboratory by changing the composition of periodically driven colloids via their size and shape distributions. These structures will maintain hyperuniformity in the small-$k$ region, while the functional form of the spectral densities away from the origin can vary widely, reflecting differences in the structures at short and intermediate lengths scales. Note the spectral density (or equivalently its corresponding direct-space two-point correlation function \cite{torquato2013random}) controls a variety of different effective properties of two-phase media, including the effective dielectric tensor \cite{rechtsman2008effective}, fluid permeability \cite{rubinstein1989flow}, mean survival time \cite{rubinstein1988diffusion}, structural-color characteristics \cite{noh2010noniridescent}, and mechanical properties \cite{xu2017microstructure}, among other quantities. Thus, the tunability of the functional
form of the spectral density and associated structures enables the fabrication of hyperuniform dispersions by self-organization that fall into the class of multifunctional materials that has been recently studied \cite{torquato2018multifunctionality}.\\

\begin{acknowledgements}
\indent The authors are grateful to Joel Lebowitz, Timothy Middlemas, JaeUk Kim, and Fang Xie for helpful discussions. This work was supported by the Air Force Office of Scientific Research Program on Mechanics of Multifunctional Materials and Microsystems under award No. FA9550-18-1-0514.
\end{acknowledgements}

\end{document}